\documentclass[]{aastex631}
\newcommand{\gtsim}{\protect\raisebox{-0.5ex}{$\:\stackrel{\textstyle >}{\sim}\:$}} 

\usepackage{color}

\shorttitle{Gravitational Redshift Detection from magnetic WD}
\shortauthors{Hayashi et al.}

\begin{document}

\title{Gravitational Redshift Detection from 
the Magnetic White Dwarf Harbored in RX\,J1712.6$-$2414}

\author{Takayuki Hayashi}
\affiliation{Center for Research and Exploration in Space Science and Technology (CRESST II), Greenbelt, MD 20771, USA}
\affiliation{Department of Physics, University of Maryland, Baltimore County, 1000 Hilltop Circle, Baltimore, MD 21250, USA}
\affiliation{NASA's Goddard Space Flight Center, X-ray Astrophysics Division, Greenbelt, MD 20771, USA}

\author{Hideyuki Mori}
\affiliation{Institude of Space and Astronautical Science (JAXA), Scientific Ballooning Research and Operation Group, Sagamihara, Kanagawa, 252-5210, Japan}


\author{Koji Mukai}
\affiliation{Center for Research and Exploration in Space Science and Technology (CRESST II), Greenbelt, MD 20771, USA}
\affiliation{Department of Physics, University of Maryland, Baltimore County, 1000 Hilltop Circle, Baltimore, MD 21250, USA}
\affiliation{NASA's Goddard Space Flight Center, X-ray Astrophysics Division, Greenbelt, MD 20771, USA}

\author{Yukikatsu Terada}
\affiliation{Institude of Space and Astronautical Science (JAXA), Department of Space Astronomy and Astrophysics, Sagamihara, Kanagawa, 252-5210, Japan}
\affiliation{Graduate School of Science and Engineering, Saitama University, 255 Shimo-Okubo, Sakura-ku, Saitama City, Saitama 338-8570, Japan}

\author{Manabu Ishida}
\affiliation{Institude of Space and Astronautical Science (JAXA), Department of Space Astronomy and Astrophysics, Sagamihara, Kanagawa, 252-5210, Japan}
\affiliation{Department of Physics, Tokyo Metropolitan University, 1-1 Minami-Osawa, Hachioji, Tokyo 192-0397, Japan}



\begin{abstract}

Gravitational redshift is a fundamental parameter 
that allows us to determine the mass-to-radius ratio 
of compact stellar objects, such as black holes, neutron stars, and white 
dwarfs (WDs).  In the X-ray spectra of the close binary system, RX\,J1712.6$-$2414, 
obtained from the {\it Chandra} High-Energy Transmission Grating observation, 
we detected significant 
redshifts for characteristic X-rays emitted from hydrogen-like magnesium, 
silicon ($\Delta E/E_{\rm rest} \sim 7 \times 10^{-4}$), and sulfur 
($\Delta E/E_{\rm rest} \sim 15 \times 10^{-4}$) ions, which are over the 
instrumental absolute energy accuracy (${\Delta E/E_{\rm rest} \sim 3.3} 
\times 10^{-4}$).  Considering 
some possible factors, such 
as Doppler shifts associated with the plasma flow, 
systemic velocity, and 
optical depth, we concluded that the major 
contributor to the observed redshift is the gravitational redshift 
of
the 
WD 
harbored in the binary system,
which is the first gravitational redshift detection from a magnetic WD.
Moreover, 
the gravitational redshift 
provides us with a new method of the WD mass measurement by invoking 
the plasma-flow theory with strong magnetic fields in close 
binaries. 
Regardless of large uncertainty, our new method estimated the WD mass 
to be $M_{\rm WD}> 0.9\,M_{\odot}$. 
\end{abstract}



\section{Introduction} \label{sec:intro}

Main-sequence (MS) stars having a mass of less than 8\,$M_{\odot}$
will evolve into a white dwarf (WD), a compact object with a radius of 
the order of $10^{4}$\,km. 
Besides, the WDs can be evolved into even by the MSs of more than 8\,$M_{\odot}$ 
in binaries
due to binary evolution.
The WD is supported against its gravity by electron degeneracy pressure.
The WD in a close binary system is fed by a mass 
accretion from its companion star. 
Hence, as the 
WD becomes more massive, 
its radius shrinks 
to 
reinforce the degeneracy pressure.
However, 
the WD mass has an upper limit (Chandrasekhar mass $\sim$ $1.38$\,$M_{\odot}$)
where the degeneracy pressure no longer supports its gravity 
\citep{1931ApJ....74...81C}.
A WD exceeding the mass limit is expected to bring about an explosion 
called a type-Ia supernova or an accretion-induced collapse into 
a neutron star.
Therefore, the mass is 
a key parameter of the WDs.
The type-Ia supernovae are used to calculate 
the distances from our galaxy 
to their host galaxies,
demonstrating 
the accelerating expansion of the Universe 
\citep{1998AJ....116.1009R,1999ApJ...517..565P}.

A gravitational redshift enables us to directly measure the 
WD mass-radius ratio.
In weak gravitational regime, 
the gravitational redshift can be written as
\begin{equation}
    v_g = cz = \frac{c\Delta E}{E_{\rm obs}} \simeq \frac{c\Delta E}{E_{\rm rest}} = 0.635\frac{M_\odot}{R_\odot}\,{\rm km\,s^{-1}},
\end{equation}
where $c$ is the speed of the light,
$z$ is the redshift parameter,
$E_{\rm obs}$ and $E_{\rm rest}$ are observed and rest-frame energies, respectively, and $\Delta E \equiv - (E_{\rm obs} - E_{\rm rest})$.
Since Einstein proposed three measurements to test the general relativity, 
one of which is to measure the gravitational redshift from 
stars \citep{1916AnP...354..769E},
the gravitational redshift has been employed to calculate 
the WD mass.
For example, the gravitational redshift of the first WD 
Sirius\,B is $v_g = cz = c\times(2.688\pm0.026)\times10^{-4} = 80.65\pm0.77$\,km\,s$^{-1}$ and thus the WD mass is $1.017\pm0.025$\,M$_{\odot}$ \citep{2018MNRAS.481.2361J}.
This technique has been applied for the WDs
in the common proper motion binaries \citep{1967ApJ...149..283G,1987ApJ...322..852K,2001AJ....121..503S},
in binaries in which the motion of both constituent star 
were well determined \citep{1998ApJ...496..449S,1999ApJ...511..916L,2006MNRAS.369.1537S,2007ApJ...667..442S,2010ApJ...715L.109V,2012MNRAS.420.3281P,2017MNRAS.470.4473P,2018MNRAS.481.2361J},
or in an open cluster \citep{1967ApJ...149..283G,1989LNP...328..378W,2001ApJ...563..987C,2019A&A...627L...8P},
where the Doppler shift can be precisely measured to 
break the degeneracy with the gravitational redshift.
However, there is no detection report of the gravitational redshift 
from highly magnetized ($B \gtrsim$ 0.1\,MG), 
spun-up ($P_{\rm spin} 
\lesssim 10^3$\,s) WDs, 
or from a WD with the X-ray.

Magnetic cataclysmic variables (mCVs)
harbors a highly-magnetized WD, which is often highly spun up 
via mass accretion.
The cataclysmic variables (CVs)
are close binaries consisting of a WD and a companion 
of late-type 
star \citep{2017PASP..129f2001M}.
In the CVs, the accreting gas from the companion
brings the angular momentum to 
the WD, causing 
the WD spin-up and frequently forming
the accretion disk around the WD.
Furthermore, as 
WD shrinks
caused by mass gain, 
its magnetic field should be strengthened \citep{2013ApJ...767L..14D}.
Such a strong magnetic field either prevents the 
accretion disk from reaching the WD surface 
(intermediate polar: IP; \citealt{1994PASP..106..209P}) or the 
prevents forming 
the accretion disk itself (polar; \citealt{1990SSRv...54..195C}).
The accreting gas is captured along the magnetic field 
and 
accelerated by the WD potential to a 
velocity 
greater than the sound 
velocity.  Hence, a standing shock 
forms near the WD surface.
According to Rankine-Hugoniot relations, the post-shock gas 
is heated up to $\sim10^{8}$\,K, 
and then the gas is highly ionized.  
While 
the post-shock gas is descending 
to the WD, 
the plasma is cooled down by emitting X-rays. 
The electrons and ions are gradually 
recombined so that the hydrogen- and helium-like (H- and He-like) ions of 
various 
elements (e.g., neon (Ne), magnesium (Mg), silicon (Si), sulfur (S), Argon 
(Ar), and iron (Fe)) are produced.
The recombined ions return 
to their ground states by emitting 
X-ray line emissions, as is indicated by observed X-ray spectra.
The ratio of the line intensities is contingent upon 
the ion population 
in the plasma flow.  
The features in the X-ray spectra, such as the
cut-off energy of the continuum 
spectra, 
line intensity ratio, and 
line energy shift caused by the Doppler shift and/or the gravitational 
redshift, allow us to measure the temperature, density, velocity, and 
gravitational potential in the post-shock plasma in principle, 
all of which are linked to the WD potential, that is, 
the WD mass.

RX\,J1712.6$-$2414, also known as 
V2400\,Oph, is an IP that
was discovered from the {\it ROSAT} All-Sky 
Survey \citep{1995MNRAS.275.1028B}.
Follow-up optical/near-infrared observation detected a circular polarization 
with the spin period of $927$\,s, indicating WD's
relatively strong magnetic field of $\bigl ( 9$--$27 \bigr ) \times 
10^{6}$\,G 
and that we always see only one of the magnetic poles; 
the accretion flow onto this magnetic pole is nearly parallel to our line of 
sight.  
This IP has been known as
a diskless IP since it 
does not show 
the $927$-s spin modulation in most cases
but shows the synodic 
$1003$-s modulation in the X-ray.
On the other hand, 
the spin modulation was observed in 2001 by {\it XMM-Newton}, and in 2005 and 2014 by 
{\it Suzaku} \citep{2019AJ....158...11J}.
However, 
the best-fit spectral model parameters of the 2001 {\it XMM-Newton} observation are consistent with those of
the 2000 {\it XMM-Newton} observation where the spin modulation 
was not detected \citep{2019AJ....158...11J},
which means the effect of whether the disk forms 
on the plasma flow structure is minor.
We observed RX\,J1712.6$-$2414 
by High-Energy Transmission Grating (HETG) of the {\it 
Chandra} observatory to 
investigate the 
velocity profile of the plasma in the accretion flow.
The HETG spectra potentially allow 
us to measure 
the Doppler shift of $\sim 30$\,km\,s$^{-1}$ \citep{2006ApJ...644L.117I}.

\section{Observation and data reduction}

We carried out the X-ray observation of RX\,J1712.6$-$2414 with 
{\it Chandra} in May 2020.
Table\,\ref{tbl:obs} shows the observation log of RX\,J1712.6$-$2414.
The observation was divided into
six intervals with the following
observation IDs: 
21274,
23038, 23039, 23244, 23267, and 23268.  The total exposure time was 
$169.15$\,ks. 
The HETG 
modules were inserted between the X-ray optics and the CCD chips.
The HETG consists of two independent gratings: the Medium Energy Grating (MEG) covers the 0.4-5.0 keV with an energy resolution $\Delta E/E \sim$ 1/300 at 2\,keV, and the High Energy Grating (HEG) covers the 0.8-10\,keV with $\Delta E/E\sim$ 200 at 6\,keV.
We chose the FAINT and Timed Exposure modes for the instrumental setup.
First, applying the latest calibration files of the instruments (the 
4.9.5 version), we reprocessed the observational data with CIAO 
v4.12 \citep{2006SPIE.6270E..1VF}.  We did not apply any filters to 
the data.  {After applying the barycentric correction, we}
then created the averaged X-ray spectra, 
following the instructions of the {\it Chandra} analysis\footnote{https://cxc.harvard.edu/ciao/}. 

\begin{table}
\caption{{\it Chandra} observations of RX\,J1712.6$-$2414}
\begin{center}
\begin{tabular}{cccc}
\hline
Observation ID & Exposure (ks) & Start date & Instrument\\\hline
21274 & 24.7 & 2020-05-30 07:31:32 & HETG$^a$ \\
23038 & 22.9 & 2020-05-29 03:35:56 & HETG$^a$ \\
23039 & 37.6 & 2020-05-06 04:05:57 & HETG$^a$ \\
23244 & 37.6 & 2020-05-06 21:24:29 & HETG$^a$ \\
23267 & 24.7 & 2020-05-31 01:29:08 & HETG$^a$ \\
3268 & 21.7 & 2020-05-31 18:53:25 & HETG$^a$ \\\hline
\multicolumn{4}{l}{$^a$High-Energy Transmission Grating}
\end{tabular}
\end{center}
\label{tbl:obs}
\end{table}%

\section{Analysis and result}
\label{sec:result}



\begin{figure}[ht]
\centering
\includegraphics[width=0.45\linewidth]{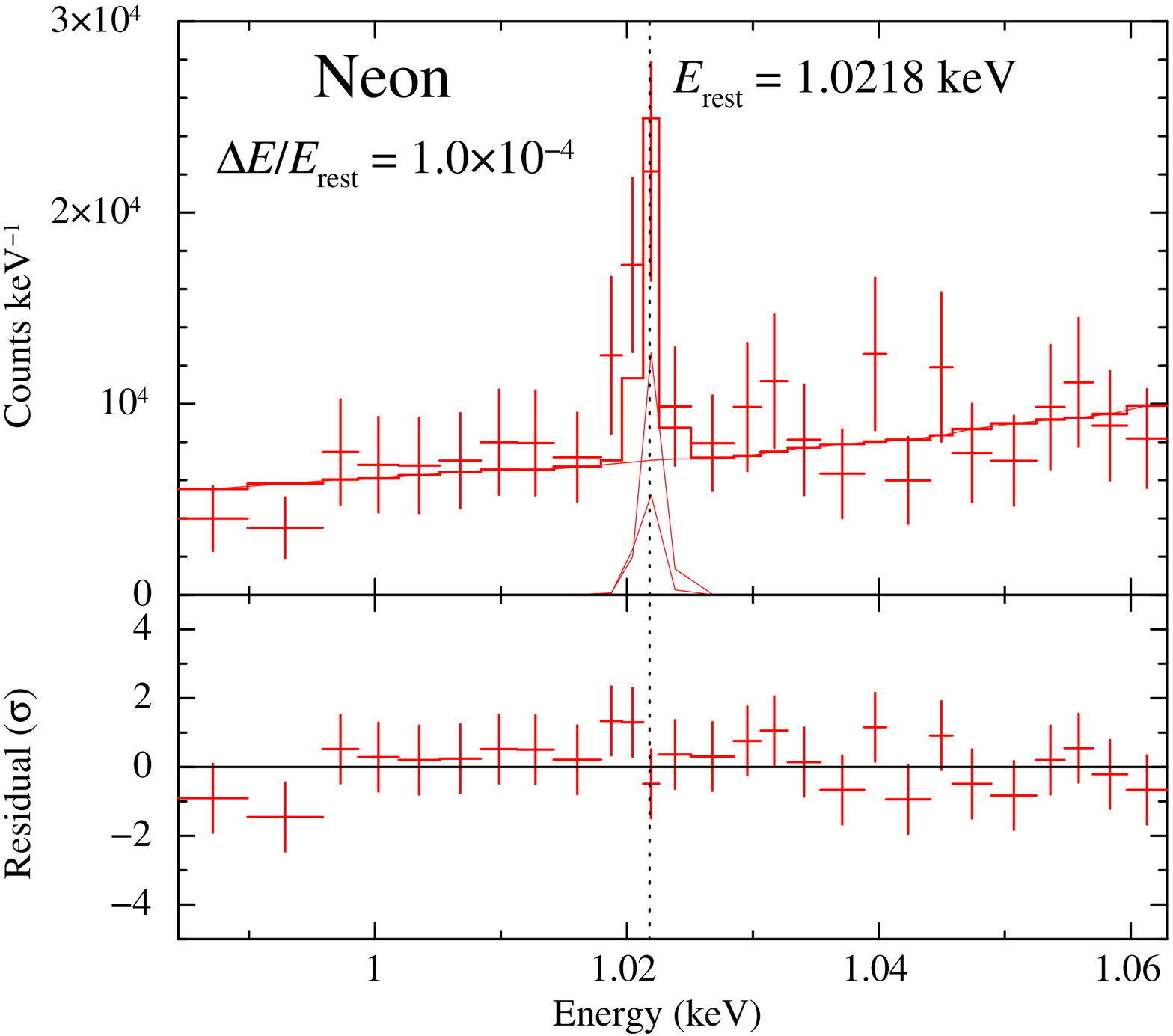}
\hspace*{2mm}
\includegraphics[width=0.45\linewidth]{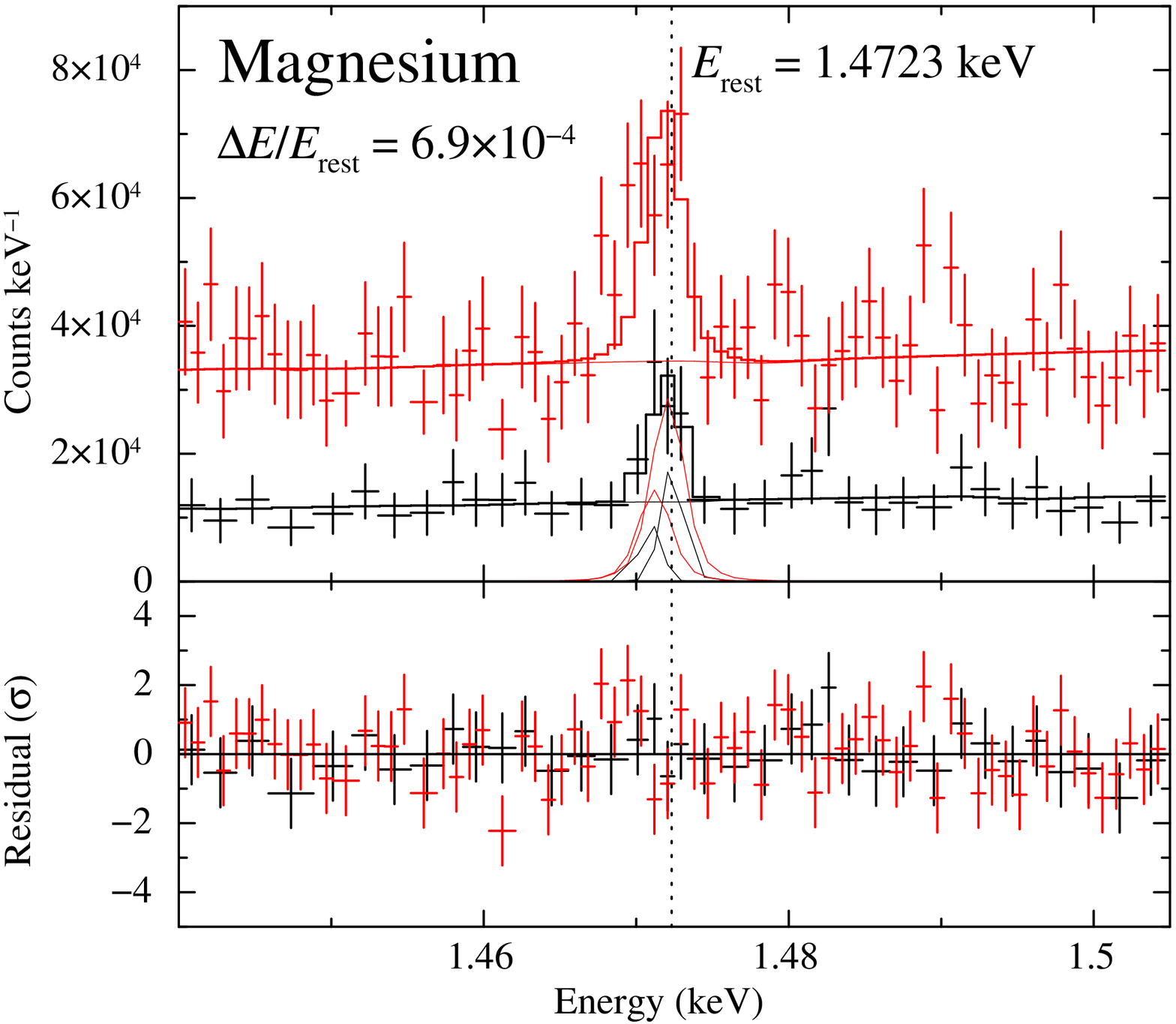} \\
\includegraphics[width=0.45\linewidth]{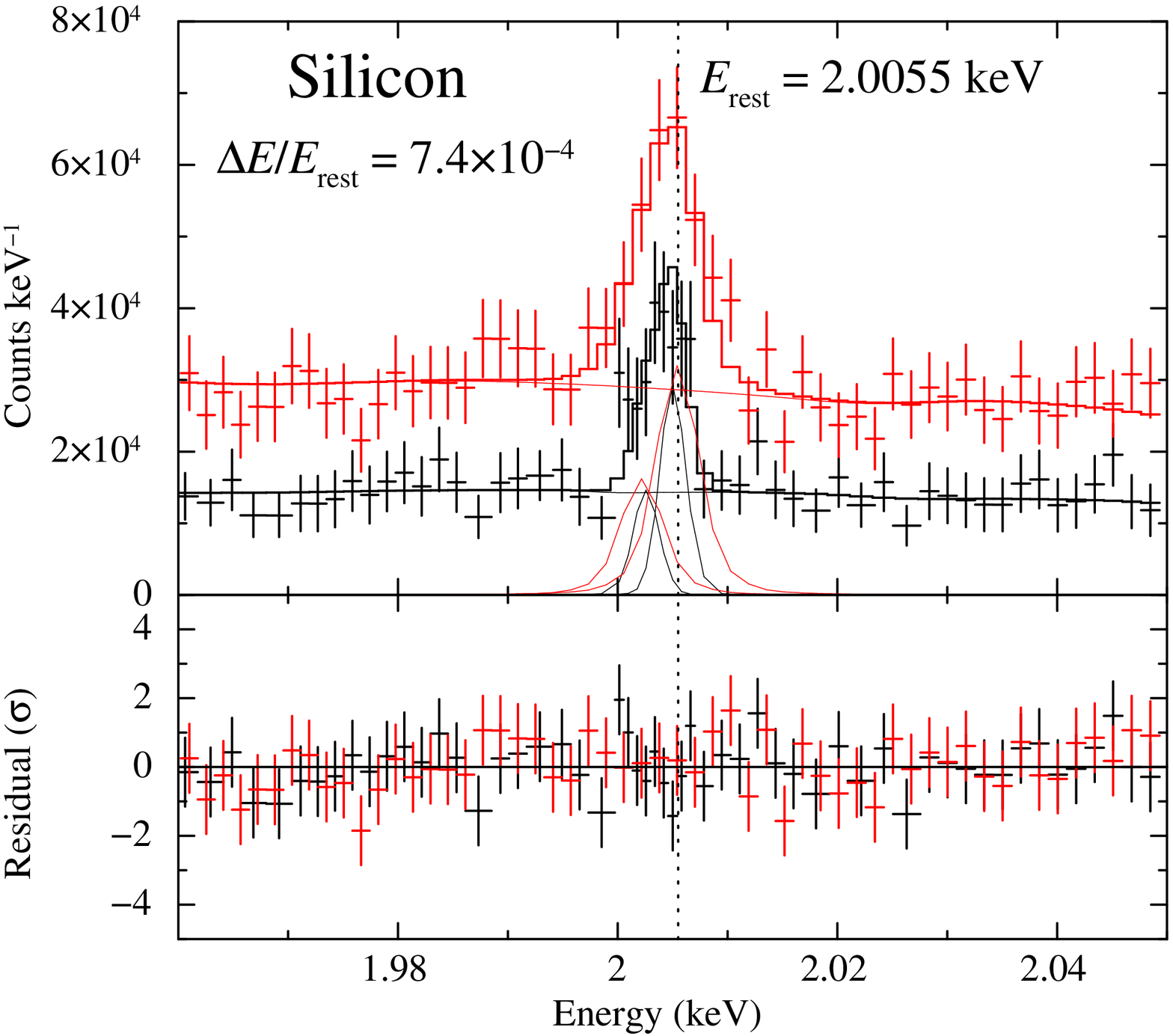}
\hspace*{2mm}
\includegraphics[width=0.45\linewidth]{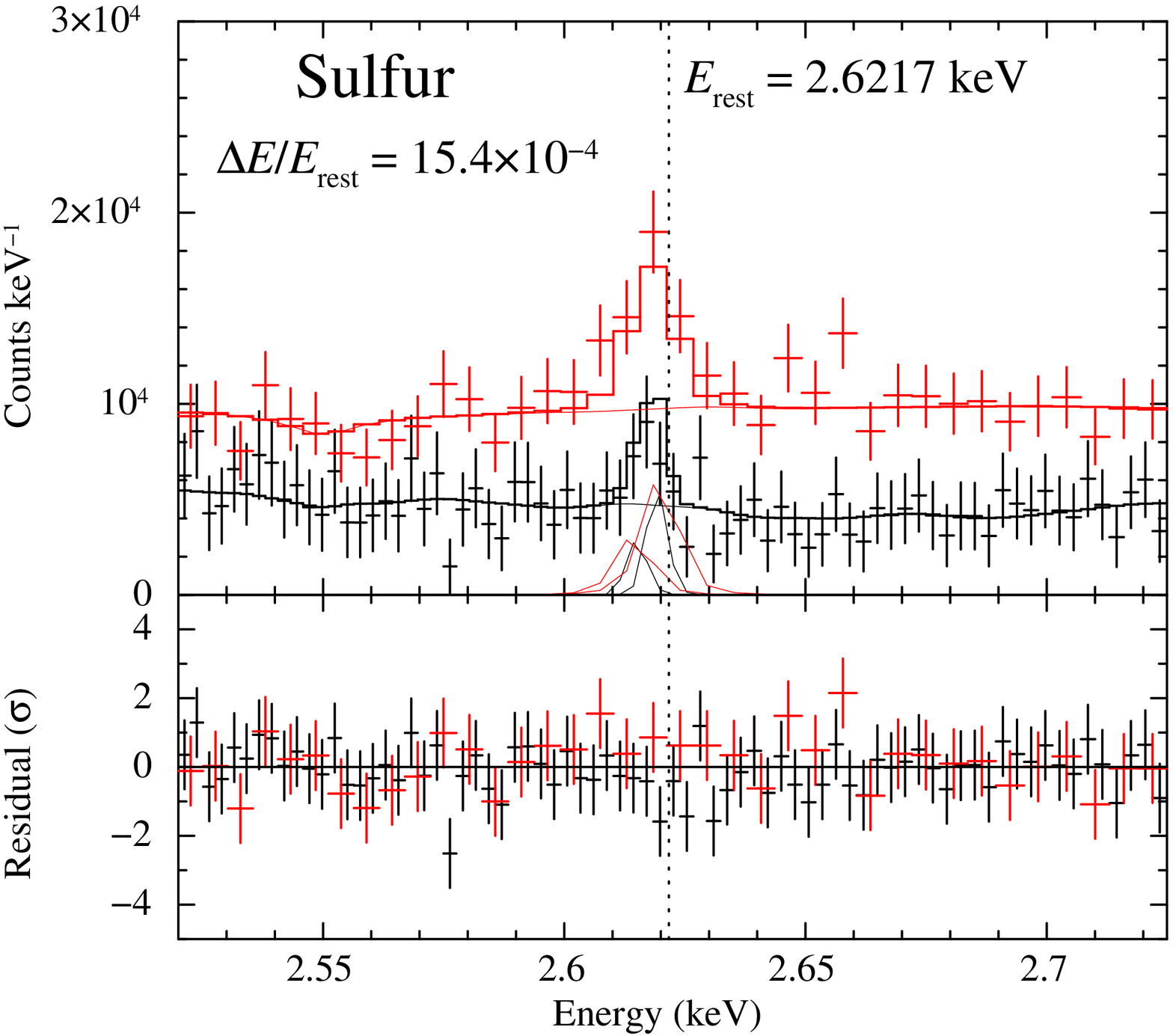} \\
\includegraphics[width=0.45\linewidth]{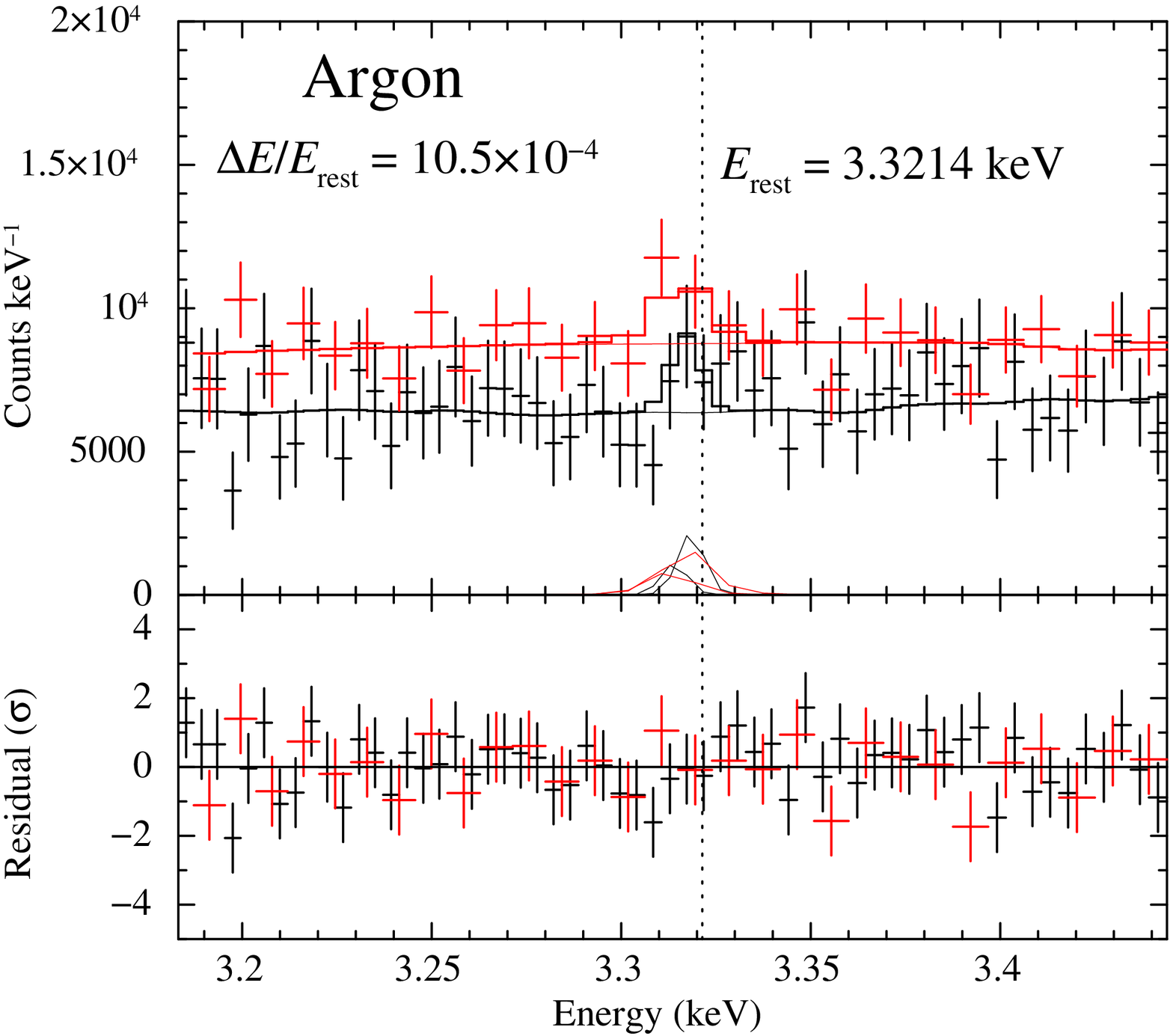}
\hspace*{2mm}
\includegraphics[width=0.45\linewidth]{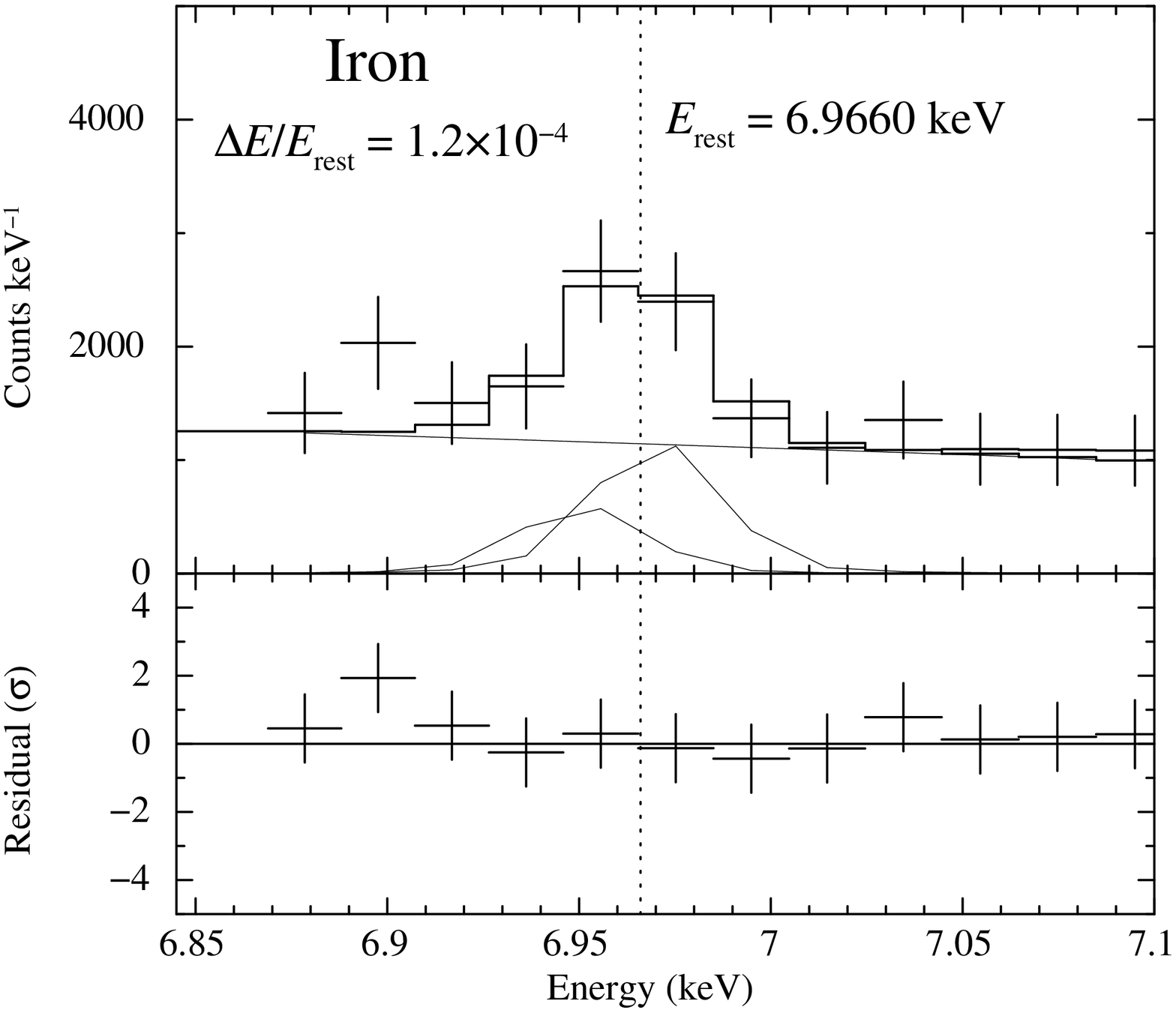} \\
\caption{
X-ray spectra obtained from the {\it Chandra} High Energy Grating (HEG; black) 
and Medium Energy Grating (MEG; red).  We show the spectra including  
emission lines from H-like ions of Ne, Mg, Si, S, Ar, and Fe.
The thick solid lines are the best-fit models.  
Meanwhile, the thin solid lines are the components of the best-fit 
models, consisting of a power-law function and two Gaussians.
The residuals from the 
best-fit model are shown at 
the bottom 
of each panel.  
We also show the position of the rest-frame energy of 
each emission line by a vertical dotted line
and the energy shift value.} 
\label{fig:Chandra_xray_spectra}
\end{figure}

Figure\,\ref{fig:Chandra_xray_spectra} shows X-ray 
HETG
spectra around 
H-like K$_\alpha$ lines of Ne, Mg, Si, S, Ar, and Fe.
We focused on 
these emission lines
to measure each energy 
shift because 
they consist of only K$_{\alpha 1}$ and K$_{\alpha 2}$, whose intensity 
ratio is invariant to the plasma temperature or density, and can be easily 
separated from lines emitted by ions in the other states.
To determine their energy centroids, 
we extracted the spectra in the energy ranges of 
$0.98$--$1.07$, 
$1.44$--$1.51$, 
$1.96$--$2.05$, 
$2.52$--$2.72$, 
$3.18$--$3.44$, 
and $6.85$--$7.10$\,keV 
for Ne, Mg, Si, S, Ar, and Fe, respectively. 
We fitted a power-law function 
to the continuum and
two Gaussians incorporating the redshift parameter 
({\tt zgauss})
to the H-like K$_{\alpha 1}$ and K$_{\alpha 2}$ 
lines by using 
Xspec (version 12.11.0; \citealt{1996ASPC..101...17A}).
The Gaussian centers were fixed at the rest-frame energies 
of K$_{\alpha 1}$ and K$_{\alpha 2}$ lines tabulated 
in Table \ref{tab:redshift} 
and their widths were fixed at 0. 
The redshift parameter
was 
common to the H-like K$_{\alpha 1}$ and K$_{\alpha 2}$ for each ions, and 
left free to fit the 
model to the data.
Their intensity ratio was fixed at the nominal of 
$I_{{\rm K}_{\alpha 1}}/I_{{\rm K}_{\alpha 2}} = 2$.
The energy ranges we selected were narrow; no photoelectric 
absorption was introduced.  

While 
the energy shifts of the K$_\alpha$ lines from 
H-like Ne, Ar, and Fe ions 
are marginal because of insufficient photon statistics, those from
H-like Mg, Si, and S ions are statistically constrained as shown 
in Table\,\ref{tab:redshift}:
$\Delta E/E_{\rm rest} = 6.9_{-0.2}^{+0.0}\times10^{-4}$
for Mg, $7.4_{-0.7}^{+0.0}\times10^{-4}$
for Si, and 
$15.4_{-4.6}^{+5.5}\times10^{-4}$
for S, which correspond to the 
line-of-sight velocities of 
$2.1_{-0.1}^{+0.0}\times10^2$, $2.2_{-0.2}^{+0.0}\times10^2$, and 
$4.6_{-1.4}^{+1.7}$\,km\,s$^{-1}\times10^2$, 
respectively.
The errors represent the statistical $90$\% confidence level.

\begin{table}[ht]
\centering
\caption{Summary of fitting to the H-like ion emission lines.}
\begin{tabular}{ccccccc}
\hline
& Ne & Mg & Si & S & Ar & Fe \\
\hline
Significance ($\sigma$) & 3.5 & 7.2 & 11.6 & 5.8 & 2.3 & 4.9 \\

$E_{\rm rest}$ of K$_{\alpha 1}$ (keV)$^{1}$ &1.0220 & 1.4726 & 2.0061 & 2.6227 & 3.3230 & 6.9732 \\
$E_{\rm rest}$ of K$_{\alpha 2}$ (keV)$^{1}$ &1.0215 & 1.4717 & 2.0043 & 2.6197 & 3.3182 & 6.9520 \\
Energy centroid of K$_{\alpha 1}$ and K$_{\alpha 2}$ (keV)$^{2}$ &1.0218 & 1.4723 & 2.0055 & 2.6217 & 3.3214 & 6.9660 \\
$z$ $\simeq\Delta E/E_{\rm rest}$ ($\times10^{-4}$)$^{2}$ & $1.0_{-\infty}^{+\infty}$ & $6.9_{-0.2}^{+0.0}$ & $7.4_{-0.7}^{+0.0}$ & $15.4_{-4.6}^{+5.5}$ & $10.5_{-12.4}^{+12.2}$ & $1.2_{-\infty}^{+\infty}$ \\
$v\,(\times10^{2}$ km\,s$^{-1}$)$^{2}$ & $0.3_{-\infty}^{+\infty}$ & $2.1_{-0.1}^{+0.0}$ & $2.2_{-0.2}^{+0.0}$ &
$4.6_{-1.4}^{+1.7}$ & $3.1\pm3.7$ & $0.4_{-\infty}^{+\infty}$ \\
Energy centroid of K$_{\alpha 1}$ and K$_{\alpha 2}$ (keV)$^{3}$ 
&&1.4722 & 2.0052 & 2.6212 \\
$z$ $\simeq\Delta E/E_{\rm rest}$ ($\times10^{-4}$)$^3$ & & 6.4$_{-0.0}^{+0.4}$ & 3.4$_{-0.4}^{+4.0}$ & 16.0$_{-6.1}^{+0.0}$\\
$v\,(\times10^{2}$ km\,s$^{-1}$)$^3$ & & $1.9_{-0.0}^{+0.1}$ & $1.0_{-0.1}^{+1.1}$ &
$4.8_{-1.8}^{+0.0}$\\
\hline
\multicolumn{7}{l}{$^{1}$ AtomDB: http://www.atomdb.org}\\
\multicolumn{7}{l}{$^{2}$ Assuming the nominal intensity ratio $I_{{\rm K}_{\alpha 1}}/I_{{\rm K}_{\alpha 2}} = 2$}\\
\multicolumn{7}{l}{$^{3}$ Assuming the optically-thick intensity ratio $I_{{\rm K}_{\alpha 1}}/I_{{\rm K}_{\alpha 2}} = 1$}
\end{tabular}
\label{tab:redshift}
\end{table}

The instrumental absolute energy accuracy
is $\Delta E/E_{\rm rest} \simeq \pm3.3\times10^{-4}$ 
(i.e., $\pm1\times10^{2}$\,km\,s$^{-1}$)
\footnote{https://cxc.harvard.edu/proposer/POG/html/index.html}. 
Even considering the instrumental accuracy, 
the emission lines demonstrate the energy shift of 
$\Delta E/E_{\rm rest} > 3 \times 10^{-4}$ and the line-of-light velocity of 
$v>1 \times 10^{2}$\,km\,s$^{-1}$.

\section{Discussion}
\label{sec:discussion}

We found redshifts of 
$z \simeq \Delta E/E_{\rm rest} = 7$ - $ 15\times10^{-4}$, 
which correspond to 
the line-of-sight velocity of 200-450\,km\,s$^{-1}$
with the H-like K$_\alpha$ lines of Mg, Si, and S from
RX\,J1712.6$-$2414.
The detected redshifts are statistically significant and
surpass the instrumental accuracy of  
 $\Delta E/E_{\rm rest}$ $\simeq$ 3 $\times 10^{-4}$
 (i.e., 100\,km\,s$^{-1}$ in the line-of-light velocity). 

\begin{figure}[ht]
\centering
\includegraphics[width=\linewidth]{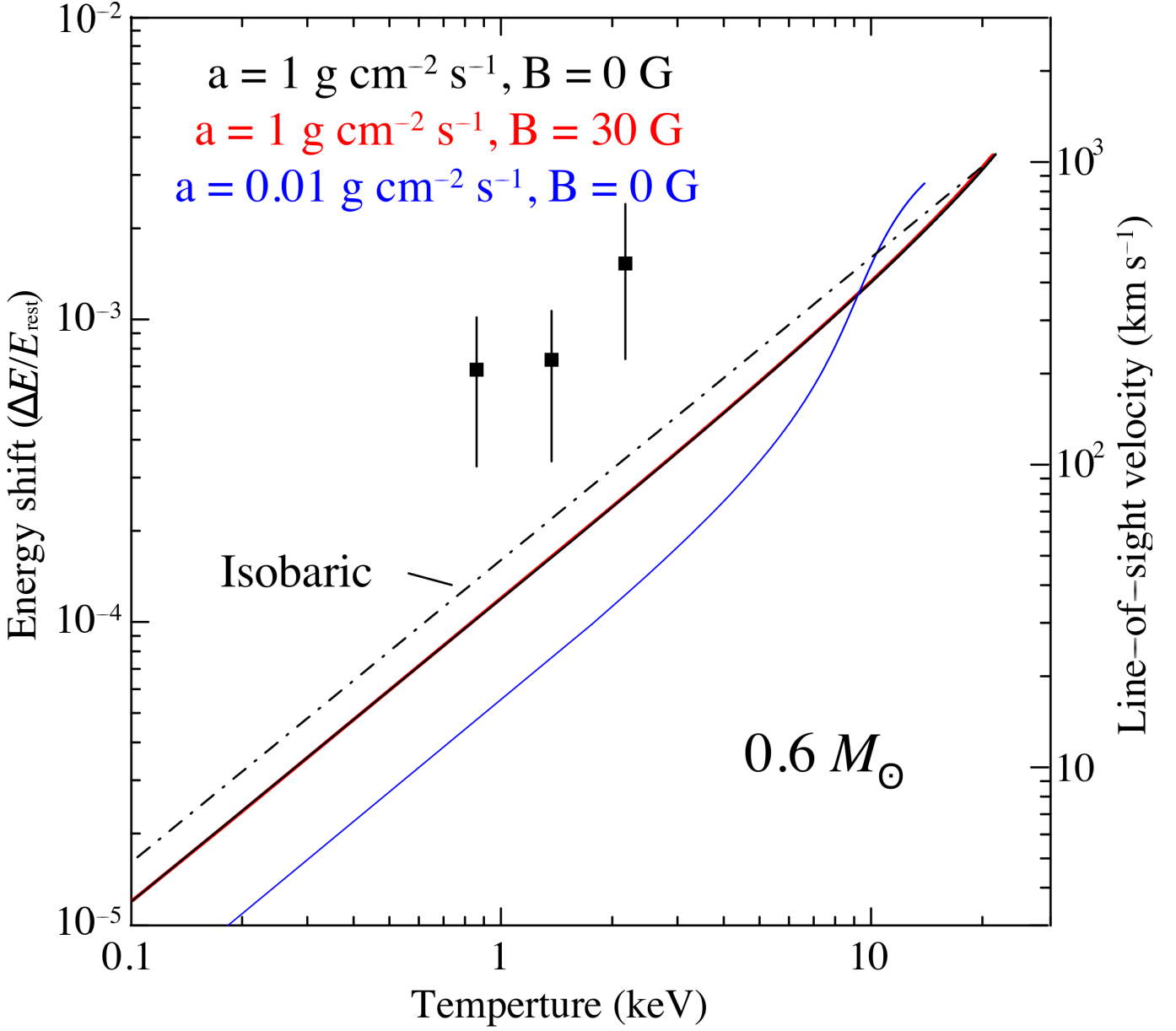}
\caption{
$\Delta E/E_{\rm rest}$ and
the line-of-sight plasma velocity
measured with the emission lines of H-like Mg, Si,
and S ions
(squares). 
The temperature of data points 
represents the line peak emissivity temperature 
of the the corresponding ion (see the text).
Each error bar shows
the sum of the 
statistical error for a $90$\% confidence
and the instrumental absolute energy uncertainty.
The lines represent theoretical temperature-velocity 
relations \citep{2014MNRAS.438.2267H} of the plasma flow 
with the WD mass of $0.6$\,$M_{\odot}$
assuming the exact pole-on geometry: thick solid lines are
the cases of 
$a$ = 1\,g\,cm$^{-2}$\,cm$^{-1}$ and 
$B$ = 0\,G (black), 
$a$ = 1\,g\,cm$^{-2}$\,cm$^{-1}$ and 
$B$ = 30M\,G (red), and
$a$ = 0.01\,g\,cm$^{-2}$\,cm$^{-1}$ and 
$B$ = 0\,G (blue). 
The black dash-dotted line shows 
the isobaric accretion flow.
}
\label{fig:T-v_0p6Msun}
\end{figure}

We realized 
that the measured redshifts cannot be explained 
by the current 
accretion models.
The magnetically channeled accretion plasma flow is modeled by 
energy and momentum equations 
combined with 
the equation of state of the ideal gas,
assuming that the plasma velocity is zero at 
the WD surface \citep{1973PThPh..50..344A}
(see Appendix).  
The solutions determine 
the temperature, density, and velocity 
profiles of the plasma
along the flow. 
The maximum temperature of the post-shock plasma in RX\,J1712.6$-$2414
was measured to be 23--26\,keV
\citep{2010A&A...520A..25Y,2016ApJ...818..136X,2019AJ....158...11J}.
Therefore, the WD mass is never less than 0.6\,$M_{\odot}$  
based on the jump conditions of the strong shock 
(equation\,\ref{eq:v0} and \ref{eq:T0}).
Figure\,\ref{fig:T-v_0p6Msun} shows 
the relations between the temperature and 
$\Delta E/E_{\rm rest}$
(i.e., the line-of-sight velocity) along the flow 
with the WD mass of 0.6\,$M_{\odot}$.
Note that we assume here the exact pole-on geometry.  Thus, 
the line-of-sight velocity equals to the actual one.
A simple analytic model with the isobaric approximation
\citep{2002apa..book.....F}
shows that the accreting plasma becomes faster with a lighter WD by comparing at a certain temperature
as is indicated by equation\,\ref{eq:Tv}. 
The plasma velocity measured 
with an emission line is 
the velocity of the local plasma whose 
temperature is at the emissivity maximum of 
the corresponding line (hereinafter called the line peak emissivity temperature).
The local plasma velocity can be measured with the centroid energy of the emission line which has the emissivity peak there.
The plasma velocity 
that the emission lines can measure
is that 
at the certain 
temperature (hereinafter called the line peak emissivity temperature),
where the corresponding species dominates the ion population.
We note that the velocity is independent of the cooling function 
(Equation\,\ref{eq:Tv}) and, therefore,
whether the cyclotron cooling is significant does not matter.
In fact, the numerical calculation
shows that the temperature$-$velocity relations 
involving and not involving the cyclotron cooling
are almost identical for 
the high specific accretion rate (i.e., accretion rate per unit 
area) of $a$ = 1\,g\,cm$^{-2}$\,cm$^{-1}$
(Figure\,\ref{fig:T-v_0p6Msun}).
Moreover, the plasma is more quickly decelerated than the prediction 
of the isobaric model because the pressure increases as the plasma descends.
A small specific accretion rate 
($a$ = 0.01\,g\,cm$^{-2}$\,cm$^{-1}$ in Figure\,\ref{fig:T-v_0p6Msun})
enlarges the increase in the pressure 
and 
makes the plasma velocity 
even slower \citep{2014MNRAS.438.2267H}.
In summary, the fastest flow is realized by the lightest 
WD mass ($0.6$\,$M_{\odot}$ for RX\,J1712.6$-$2414) 
and a high enough specific accretion rate. 
The observed line-of-sight velocities of 
$\gtsim 1\times10^{2}$\,km\,s$^{-1}$ 
are significantly faster 
than the theoretical fastest flow
($\simeq$ 30\,km\,s$^{-1}$ and 80\,km\,s$^{-1}$
at the line peak emissivity temperatures of Mg and S, 
respectively).

We investigated other possibilities that increase the
$\Delta E/E_{\rm rest}$. 
RX\,J1712.6$-$2414 is located at ($l$, $b$) = 
($+359^{\circ}\!\!.87$, $+8^{\circ}\!\!.74$)
in the Galactic coordinate,  
and its distance and proper motions ($\mu_{\rm ra}$, $\mu_{\rm dec}$) 
are 699.7$_{-10.8}^{+9.7}$\,pc and (-1.765, 2.557) in mas\,yr$^{-1}$ \citep{refId0,2021AJ....161..147B}, respectively.
Assuming the space motion along the galactic pole $W=0$,
the radial velocity is calculated at $-$2\,km\,s$^{-1}$.
Indeed, the optical spectra obtained from the South African Astronomical 
Observatory revealed that the systemic velocity is less than 
$20$\,km\,s$^{-1}$\citep{1995MNRAS.275.1028B}.  
Furthermore, the line-of-sight velocity 
associated with the binary motion 
is nullified 
in our phase-averaged spectra
and does not affect the result.

Although significant optical depth 
shifts the energy centroids of the emission lines 
\citep{2002A&A...385..968D},
it is not enough to explain the observed redshifts. 
The K$_\alpha$ lines are ensembles of 
the K$_{\alpha2}$ and K$_{\alpha1}$ lines,
whose intensity ratio ($I_{{\rm K}_{\alpha2}}/I_{{\rm K}_{\alpha1}}$) 
affects the energy centroid of the K$_\alpha$ lines. 
The optical depth at the K$_{\alpha1}$ line
is double that at the K$_{\alpha2}$ line. 
Therefore, the K$_{\alpha1}$ line is more easily attenuated 
than the K$_{\alpha2}$ line; thus, 
the K$_{\alpha}$ line energy centroid shifts toward the red side. 
This effect approximately halves 
$I_{{\rm K}_{\alpha1}}/I_{{\rm K}_{\alpha2}}$
and makes it unity at the optically thick limit \citep{1990JQSRT..44..275K,1999A&A...351L..23M} (see Appendix). 
We fitted the power-law and 2-Gaussians model in the same manner in 
$\S$\ref{sec:result} 
by 
assuming
$I_{{\rm K}_{\alpha1}}/I_{{\rm K}_{\alpha2}}=1$.
The computed $\Delta E/E_{\rm rest}$ and velocity 
are also presented 
in Table\,\ref{tab:redshift},
which still shows the velocities of $\gtsim 1\times10^{2}$\,km\,s$^{-1}$.
Although all of them are consistent with the corresponding result with 
$I_{{\rm K}_{\alpha1}}/I_{{\rm K}_{\alpha2}}=2$ within the statistical 
error, the H-like Si K$_\alpha$ line gave us the best-fit $\Delta 
E/E_{\rm rest}$, different from the corresponding by $4\times10^{-4}$.
However, this line showed 
a fine structure in the fitting residual (see 
Figure\,\ref{fig:opt_thick_fit}), 
implying 
that optically thick limit (i.e., 
$I_{{\rm K}_{\alpha2}}/I_{{\rm K}_{\alpha1}}=1$) is not a good assumption.

The H-like K$_\alpha$ lines 
from the pre-shock accreting gas do not contaminate 
those lines from the post-shock plasma.
The pre-shock accreting plasma close to the shock is 
photoionized by the X-ray irradiation 
from the post-shock plasma 
and emits H-like K$_\alpha$ lines \citep{2010ApJ...711.1333L}.
However, the velocity of the pre-shock gas is 4.1$\times10^{3}$\,km\,s$^{-1}$ and $\Delta E/E_{\rm rest}$ = 1.4$\times10^{-2}$ with the pole-on geometry even if the WD is light as much as possible i.e., 0.6\,$M_{\odot}$.
Such highly Doppler-shifted line spectroscopically
goes out of the line of the same ion from
the post-shock plasma. 

\begin{figure}[ht]
\centering
\includegraphics[width=\linewidth]{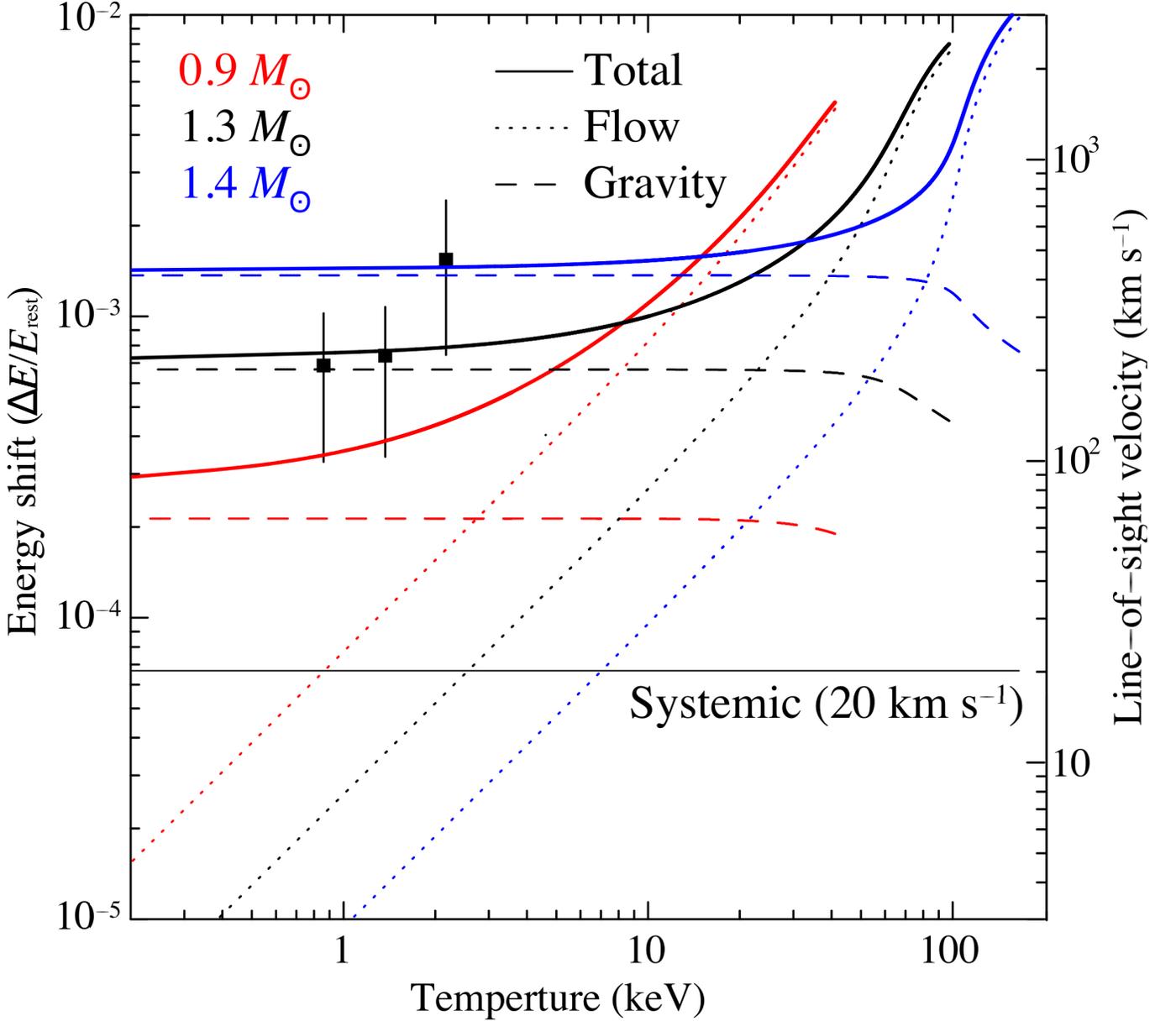}
\caption{
Same as Figure\,\ref{fig:T-v_0p6Msun}, except for
theoretical calculations involving 
the binary system motion and the gravitational redshift. 
Thick solid lines represent the
calculations with 
the WD mass of $0.9\,M_{\odot}$ (red), $1.3\,M_{\odot}$ (black), and $1.4\,M_{\odot}$ (blue).  
The dashes, dotted, and thin lines 
show the components of the gravitational redshift,
the plasma flow velocity \citep{2014MNRAS.438.2267H}, and the binary
systemic velocity, respectively.
The pole-on geometry is assumed for the calculations.
}
\label{fig:T-v_relation}
\end{figure}

Consequently, we 
conclude 
that the measured $\Delta E/E_{\rm rest}$ requires 
a gravitational redshift 
caused by the WD. 
Considering 
the
gravitational redshift to our 
accretion-flow model \citep{2014MNRAS.438.2267H},
as well as the systemic 
velocity and the instrumental absolute energy accuracy,
we estimated the WD mass of RX\,J1712.6$-$2414 to be 
$> 0.9$\,$M_{\odot}$ (Figure\,\ref{fig:T-v_relation}).
The WD mass estimated by previous works (e.g., $0.62^{+0.06}_{-0.05}\,M_{\odot}$\citep{2010A&A...520A..25Y},
$0.72\pm0.05\,M_{\odot}$\citep{2019MNRAS.482.3622S},
and $0.67^{+0.06}_{-0.05}\,M_{\odot}$\citep{2020MNRAS.498.3457S}) 
is lighter than ours. 
One plausible explanation
for the discrepancy is
the cyclotron cooling 
that softens the X-ray spectrum 
and was not considered in the previous mass estimations. 
The magnetic field of RX\,J1712.6$-$2414 is $(9-27)\times10^6$\,G 
and comparable to that of polars 
in which the cyclotron cooling is significant \citep{1994ApJ...426..664W}.
Another reasonable reason is
the X-ray reflection that is maximized at the pole-on geometry
and causes 
complicated systematic error \citep{2021MNRAS.504.3651H}.
In previous studies, X-ray reflection was not taken into account.

Lastly, we note that more precise spectral modeling 
reduces 
the contribution of the plasma velocity 
to the redshift, which 
improves the accuracy of
the gravitational redshift estimation. 
We assume that the H-like K$_\alpha$ lines are emitted at the corresponding line peak emissivity temperature.
However, the lines are indeed emitted from a temperature range specified for each ion.
Accreting plasma neighboring on 
the WD has higher density
and thus 
emits more intense X-rays.
Meanwhile, the neighboring plasma
is slower so that the velocity average
weighted over the X-ray intensity 
is slower than the velocity at the line peak emissivity temperature we used in Figures\,\ref{fig:T-v_0p6Msun} and \ref{fig:T-v_relation}.
The more precise spectral modeling may 
necessitate greater 
WD mass,
but it is beyond our main aim to report the gravitational redshift detection.

\section{Conclusion}

We observed the diskless intermediate polar RX\,J1712.6-2414 
with the High-Energy Transmission Grating (HETG) of 
the Chandra Observatory to study
the velocity profile of the plasma in the accretion flow.
We found significant redshifts for the K$\alpha$ lines 
of hydrogen-like magnesium, silicon 
($\Delta E/E_{\rm rest} \sim 7 \times 10^{-4}$), and sulfur 
($\Delta E/E_{\rm rest} \sim 15 \times 10^{-4}$) ions, 
which are above the instrumental absolute energy accuracy 
(${\Delta E/E_{\rm rest} \sim 3.3} \times 10^{-4}$).  
We considered several factors producing the redshift,
such as
the Doppler shift associated with the plasma flow velocity and the systemic velocity,
the optical depth, and the gravitational redshift,
and then concluded that the gravitational redshift is the
major contributor to the observed redshift.
This is the first gravitational redshift detection from a magnetic WD.
The gravitational redshift 
provides us with a new method of the WD mass measurement,
which estimates the WD mass to be $M_{\rm WD}> 0.9\,M_{\odot}$.

\appendix 

\subsection*{Plasma flow model}
\label{sec:model}
An accretion plasma flow channelled 
by a magnetic field is modeled 
as a 1-dimensional flow.
Fundamental equations of a 1-dimensional flow are
the mass continuity equation:
\begin{eqnarray}
\frac{{\rm d}}{{\rm d}z}(\rho v) = 0\label{eq:conti},
\end{eqnarray}
the momentum equation:
\begin{eqnarray}
\rho v \frac{{\rm d}v}{{\rm d}z}+\frac{{\rm d}P}{{\rm d}z} = \rho F, 
\label{eq:momentum}
\end{eqnarray}
and the energy equation:
\begin{eqnarray}
\frac{\rm d}{{\rm d}z}\left[v\left(\frac{1}{2}\rho v^2+\frac{\gamma P}{\gamma-1}\right)\right] = \rho v F-\varepsilon.
\label{eq:energy}
\end{eqnarray}
Here,
$\rho$ denotes 
mass density,
$v$ presents 
flow velocity,
$P$ denotes 
pressure,
$F$ is an external force,
$\epsilon$ is a radiative cooling rate, 
and $\gamma$ is an adiabatic index of $5/3$.
The integral form of 
equation\,\ref{eq:conti} is
\begin{eqnarray}
\rho v = a\label{eq:conti_integ},
\end{eqnarray}
where $a$ is called a specific accretion rate, 
that is,
the accretion rate per unit area.
The simultaneous equations 
\ref{eq:momentum}, \ref{eq:energy} and \ref{eq:conti_integ}
are resolved
under initial conditions 
derived from the strong-shock jump condition 
calculated by the Rankine-Hugoniot relations with the free-fall velocity:
\begin{eqnarray}
v_0 &=& 0.25\sqrt{2GM_{\rm WD}/(R_{\rm WD} + h)}, \label{eq:v0}\\ 
\rho_0 &=& \frac{a}{v_0},\\ 
P_0 &=& 3av_0 \label{eq:P0},\\
T_0 &=& 3\frac{\mu m_{\rm H}}{k}v_0^2 \label{eq:T0},
\label{eq:3initial_eqi_cropper}
\end{eqnarray}
where $M_{\rm WD}$ is a WD mass, $R_{\rm WD}$ is a WD radius, $G$ is the constant of gravitation, and $h$ is a shock height.
The equation of state for the ideal gas is used:
\begin{eqnarray}
P = \frac{\rho k T}{\mu m_{\rm H}} \label{eq:eos}.
\end{eqnarray}
A boundary condition of soft landing
is also assumed at the WD surface:
\begin{eqnarray}
v_{\rm WD} = 0.  \label{eq:softlanding}
\end{eqnarray}

By assuming that the shock height is negligibly 
low compared with the WD radius ($R_{\rm WD} \gg h$, i.e., the post-shock region is low enough), 
the Bernoulli's principle requires
the total of the pressure and the ram pressure is constant:
\begin{eqnarray}
P_0 + \rho_0 v_0^2 = P_{\rm WD} = {\rm constant}, \label{eq:Bernoulli}
\end{eqnarray}
where $P_{\rm WD}$ is the pressure at the WD surface
at which the velocity is zero.
From equations \ref{eq:conti_integ} and \ref{eq:P0},
we obtain
\begin{eqnarray}
P_0 = 3\rho_0 v_0^2.
\end{eqnarray}
In other words, the pressure is increased only by a factor of 4/3 
through the entire flow.
Thus, an isobaric flow is a good approximation. 
With the constant pressure, from equations \ref{eq:conti_integ} and \ref{eq:eos}, the following is derived:
\begin{eqnarray}
\frac{v}{v_0} = \frac{\rho_0}{\rho} = \frac{T}{T_0}.\label{v-rho-T}
\end{eqnarray}
From equations \ref{eq:v0}, \ref{eq:T0} and \ref{v-rho-T},
the relation between the temperature and the velocity 
is written as
\begin{eqnarray}
v = \frac{v_0}{T_0}T = \left(\frac{3\mu m_{\rm H}}{4k}\sqrt{\frac{2G M_{\rm WD}}{R_{\rm WD}}}\right)^{-1}T.\label{eq:Tv}
\end{eqnarray}
A result of this simple analytical model
is shown in the dot-dash line (labelled with ``isobaric'') in Figures\,
\ref{fig:T-v_0p6Msun} and \ref{fig:T-v_relation}.

Assuming only the Bremsstrahlung 
for the radiation cooling process,
equations \ref{eq:momentum}, \ref{eq:energy}, \ref{eq:conti_integ},
and \ref{v-rho-T}
give us a simple thermal function of $z$: 
\begin{eqnarray}
\frac{T}{T_0} =\left(\frac{z}{h}\right)^{2/5},
\end{eqnarray}
where
\begin{eqnarray}
h = kT_0^{1/2}v_0/({\mu m_{\rm H}} \Lambda_{\rm m,br} \rho_0)
\end{eqnarray}
and
\begin{eqnarray}
\Lambda_{\rm m,br} \sim 7\times10^{20}\,{\rm erg\,g^{-1}\,s^{-1}}.
\end{eqnarray}

More realistic models have been numerically calculated.
The cooling via line emission is included into the cooling function according to plasma codes \citep{2010A&A...520A..25Y,2014MNRAS.438.2267H}. 
Some models involve the cyclotron cooling, 
which is usually important in the 
polars
(this radiation mainly appears in the infrared band), 
as well by assuming 
optically thick radiation \citep{1980PThPh..64.1986W,1999MNRAS.306..684C,2021ApJS..256...45B}.
Note that the difference in the cooling function does not affect the
relation between the temperature and velocity, as shown in equation\,\ref{eq:Tv}.
Moreover, in a real plasma flow with finite height, 
the gravitational force works.
$F$ in equations \ref{eq:momentum} and \ref{eq:energy}
is represented as follows to include this effect.
\begin{eqnarray}
 F = \frac{G M_{\rm WD}}{(R_{\rm WD}+z)^2}.
\end{eqnarray}
The calculation of the realistic model are shown 
in Figures\,
\ref{fig:T-v_0p6Msun} and \ref{fig:T-v_relation}. 

\subsection*{Optical depth}
\begin{figure}[ht]
\centering
\includegraphics[width=\linewidth]{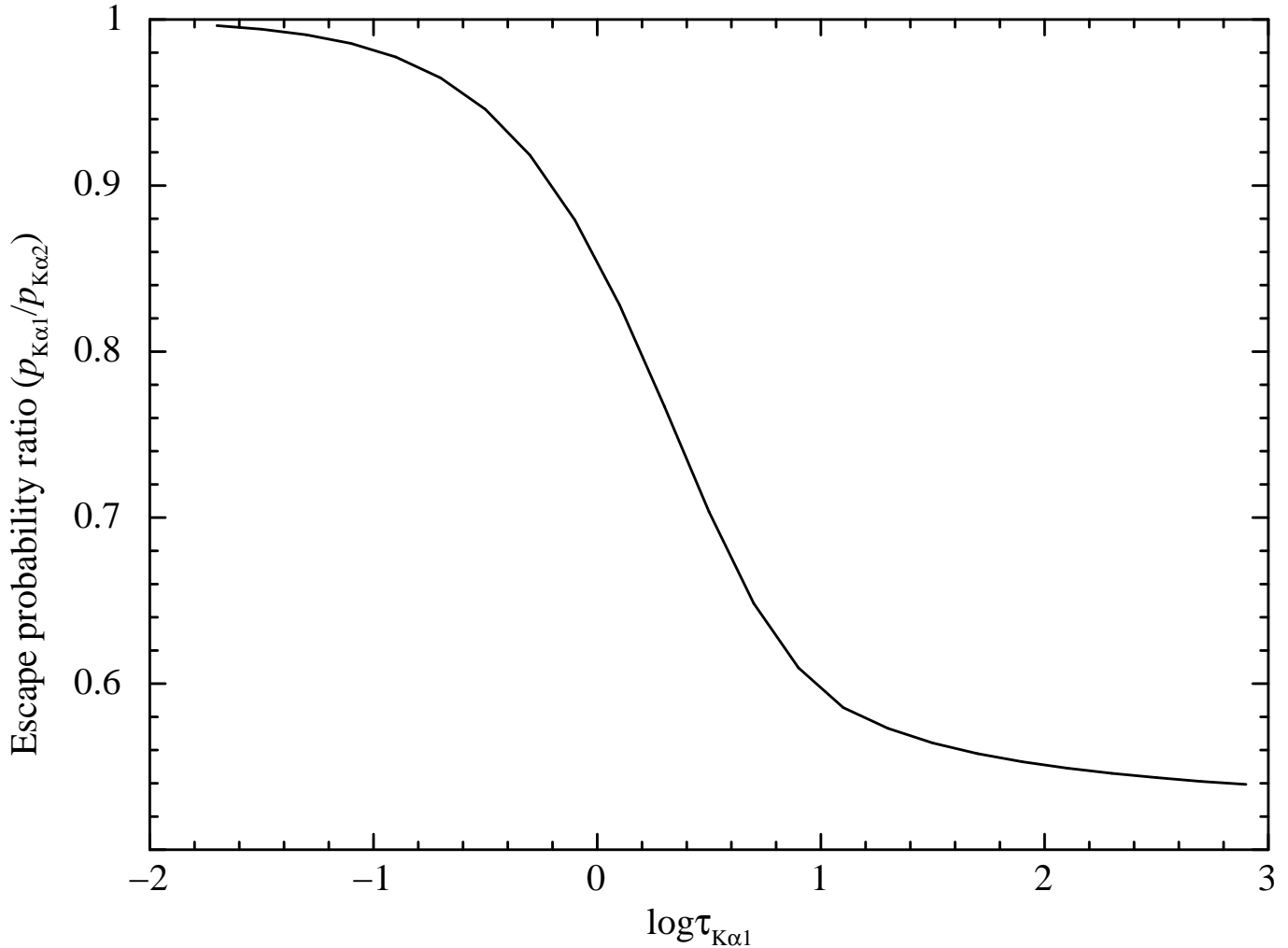}
\caption{
The ratio of the escape probability of 
K$_\alpha1$ line to K$_\alpha2$ line
as a function of optical depth of the K$_\alpha1$ line (log$\tau_{\rm K_{\alpha1}}$).
}
\label{fig:p_escape_ratio}
\end{figure}

The observed intensity ratio of K$_{\alpha1}$ and K$_{\alpha2}$
is represented by
\begin{eqnarray}
\frac{I_{{\rm K}_{\alpha 1}}}{I_{{\rm K}_{\alpha 2}}} =
\frac{I_{{\rm 0,K}_{\alpha 1}}}{I_{{\rm 0,K}_{\alpha 2}}}
\frac{p_{{\rm K}_{\alpha 1}}}{p_{{\rm K}_{\alpha 2}}} =
2\frac{p_{{\rm K}_{\alpha 1}}}{p_{{\rm K}_{\alpha 2}}}\label{eq:obs_i_ratio}
\end{eqnarray}
where $I_0$ is the nominal intensity, that is,
the intensity at the moment of emission,
and $p$ is 
the probability that an X-ray photon escapes out 
from the post-shock plasma.
The optical depth ($\tau$) at the central energy of a line
is proportional to the oscillator strength ($f$)
of the corresponding transition \citep{1978pas..conf.1453J}.
Moreover, the oscillator strength of a K$_{\alpha1}$ line
is double that of 
the K$_{\alpha2}$ line as
\begin{eqnarray}
\frac{\tau_{{\rm K}_{\alpha 1}}}{\tau_{{\rm K}_{\alpha 2}}} =
\frac{f_{{\rm K}_{\alpha 1}}}{f_{{\rm K}_{\alpha 2}}} = 2.
\end{eqnarray}
The ratio of the escape probability 
was calculated \citep{1990JQSRT..44..275K} for a situation
in which the emitters (i.e., excited H-like ions) and 
absorbers (i.e., H-like ions at the ground state)
are identically distributed as in the post-shock plasma.
Figure\,\ref{fig:p_escape_ratio} depicts 
the escape probability ratio 
for lines with an optical depth ratio of 2,
i.e.,
K$_{\alpha 1}$ and K$_{\alpha 2}$.
At the optically thick limit,
the escape probability ratio approximately converges on 0.5
and 
$I_{{\rm K}_{\alpha 1}}/I_{{\rm K}_{\alpha 2}} \simeq 1$
at the optically thick limit 
from equation\,\ref{eq:obs_i_ratio}.

\begin{figure}[ht]
\centering
\includegraphics[width=0.45\linewidth]{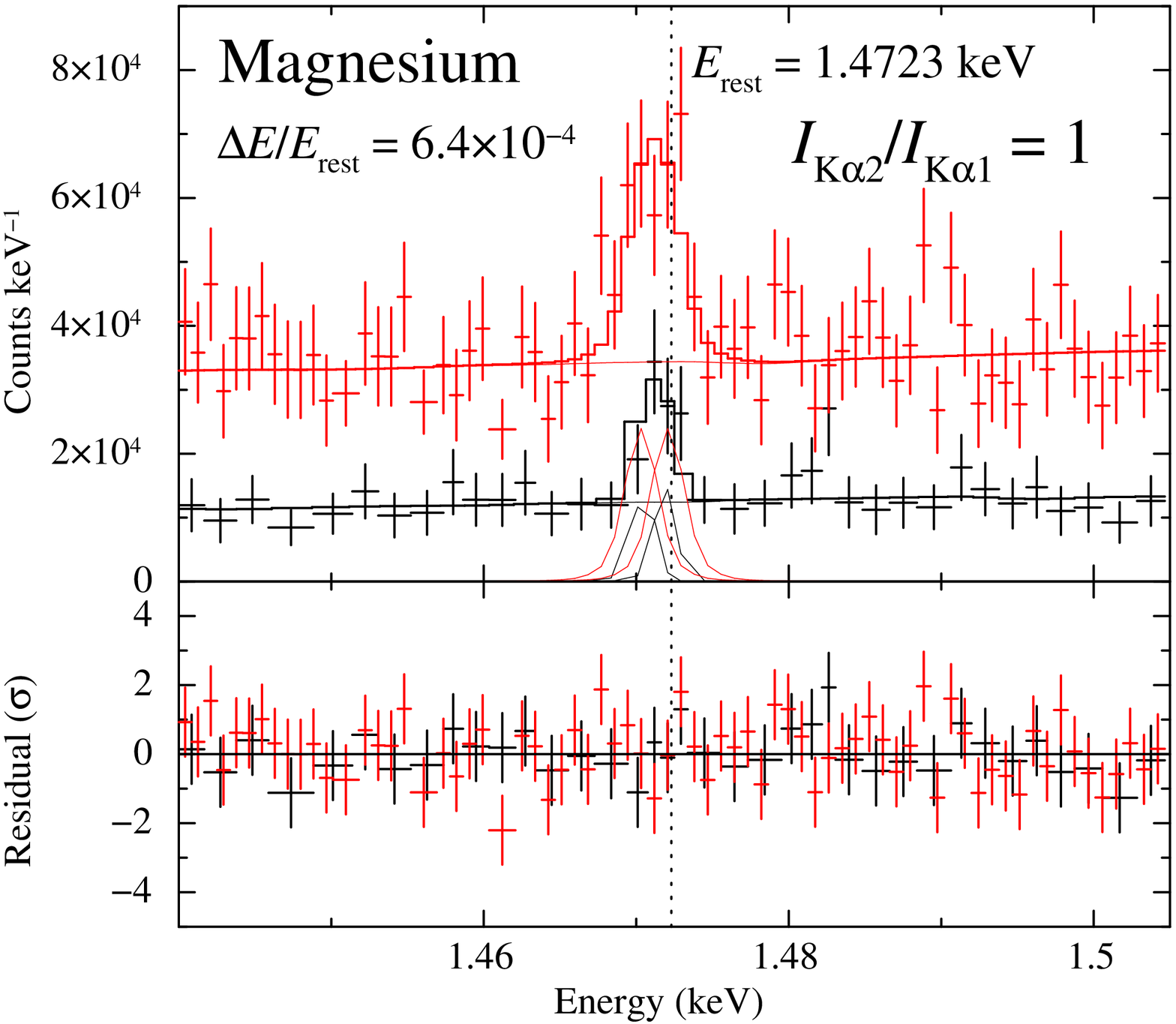}
\hspace*{2mm}
\includegraphics[width=0.45\linewidth]{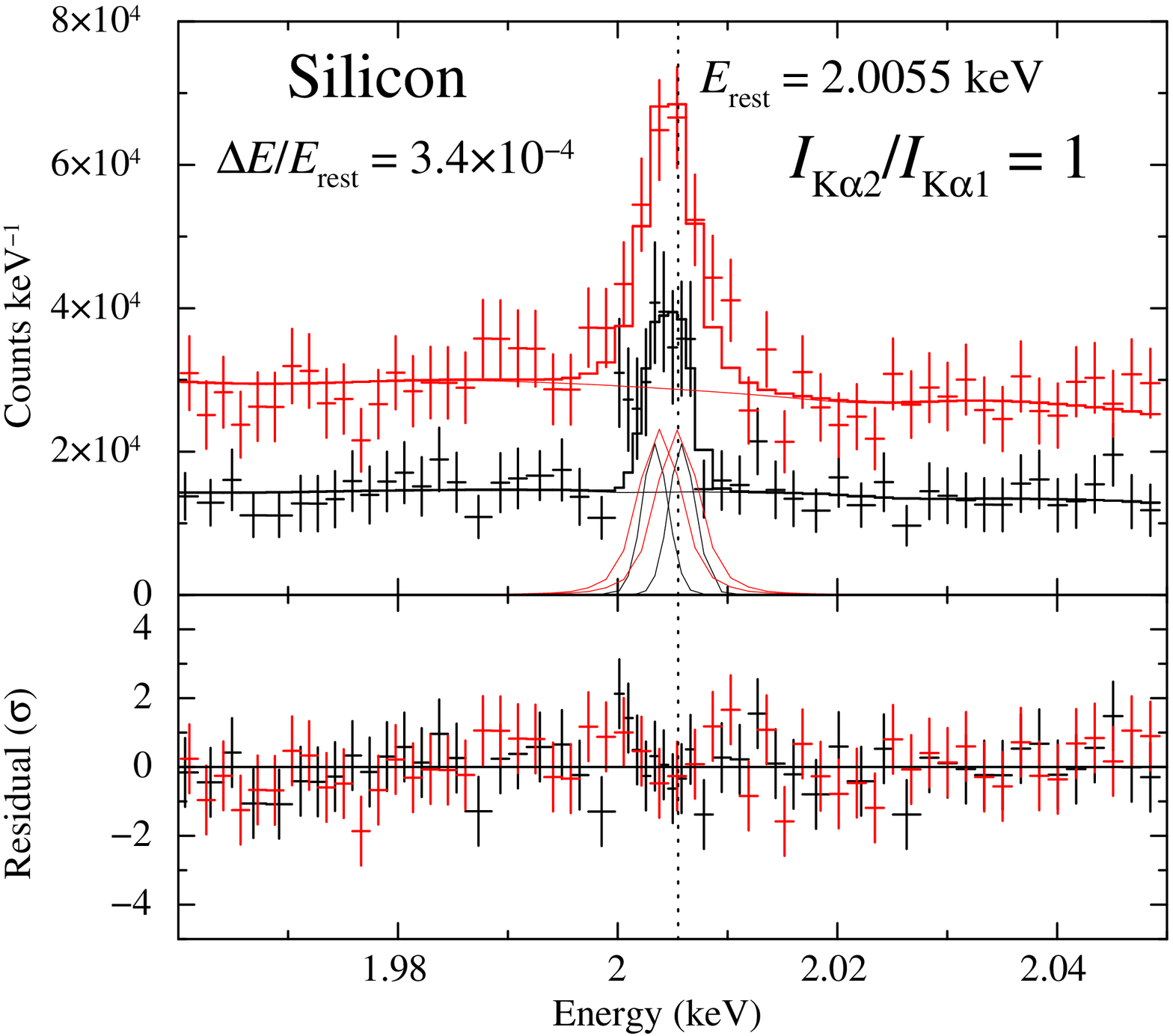}\\
\includegraphics[width=0.45\linewidth]{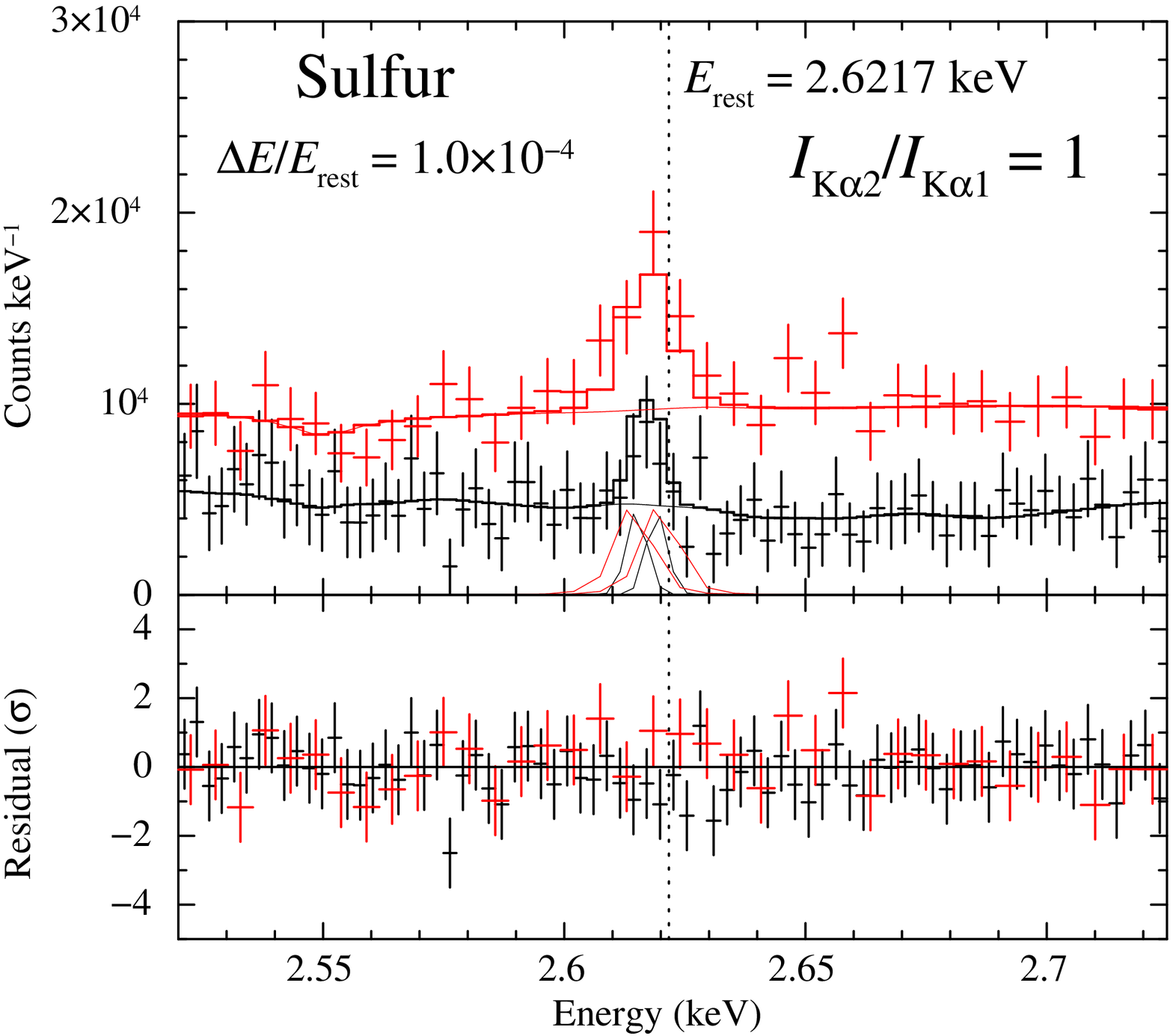}
\caption{
Same as the spectra of Mg, Si, and S in 
Figure\,\ref{fig:Chandra_xray_spectra}, except for
the optically-thick limit assumption ($I_{{\rm K}_{\alpha 1}}/I_{{\rm K}_{\alpha 2}}$ = 1 for H-like ion K$_\alpha$ lines).
}
\label{fig:opt_thick_fit}
\end{figure}

We fitted the power-law and 2-Gaussians model for Mg, Si, and S
in the same manner in the $\S$\ref{sec:result}
by using $I_{{\rm K}_{\alpha1}}/I_{{\rm K}_{\alpha2}}=1$.
Figure\,\ref{fig:opt_thick_fit}
shows the best-fit models with the data
and Table\,\ref{tab:redshift}
shows the best-fit energy shift and velocity.

\section*{Acknowledgement}
The authors are grateful to all of the {\it Chandra}
project members for developing the instruments and their software, the spacecraft operations, and the calibrations. 
We also thank Maruzen-Yushodo Co. Ltd. and Xtra Inc. for their language editing service of our English.
This research has made use of data obtained from the Chandra Data Archive and the Chandra Source Catalog, and software provided by the Chandra X-ray Center (CXC) in the application packages CIAO and Sherpa.
Support for this work was provided 
by the National Aeronautics and Space Administration 
through Chandra Award Number GO9-20022A issued by the Chandra X-ray Center, which is operated by the Smithsonian Astrophysical Observatory for and on behalf of the National Aeronautics Space Administration under contract NAS8-03060.
This work was also supported by JSPS Grant-in-Aid for Scientific Research(C) Grant Number JP21K03623.



\bibliography{sample631}{}
\bibliographystyle{aasjournal}



\end{document}